# Enhanced single photon emission from positioned InP/GaInP quantum dots coupled to a confined Tamm-plasmon mode

T. Braun[1], V. Baumann[1], O. Iff[1], S. Höfling[1,2], C. Schneider[1], and M. Kamp[1]

[1]Technische Physik, Physikalisches Institut and Wilhelm Conrad Röntgen-Research Center for Complex Material Systems, Universität Würzburg, Am Hubland, D-97074, Würzburg, Germany

[2]SUPA, School of Physics and Astronomy, University of St Andrews, St Andrews, KY16 9SS, United Kingdom

Abstract: We report on the enhancement of the spontaneous emission in the visible red spectral range from site-controlled InP/GaInP quantum dots by resonant coupling to Tamm-plasmon modes confined beneath gold disks in a hybrid metal/semiconductor structure. The enhancement of the emission intensity is confirmed by spatially resolved micro-photoluminescence area scans and temperature dependent measurements. Single photon emission from our coupled system is verified via second order autocorrelation measurements. We observe bright single quantum dot emission of up to ~173000 detected photons per second at a repetition rate of the excitation source of 82 MHz, and calculate an extraction efficiency of our device as high as 7%.

The fabrication of efficient single photon sources (SPSs) is a demanding task in the field of quantum communication[1]. Especially quantum dots (QDs) are promising candidates for applications as quantum light sources and have been investigated intensively concerning their capability to emit single[1,2,3], indistinguishable[4,5] and entangled photons[6,7] on demand. However, for the realization of high-performance SPSs an efficient extraction of the photons is required. Past research activities mainly focused on the integration of QDs in structures such as photonic cavities or waveguides in order to enhance the efficiency of the spontaneous emission, or suppress the emission into leaky modes[7,8,9,10,11]. Recently enhancement as well as inhibition of the spontaneous emission from self-assembled In(Ga)As/GaAs QDs coupled to confined Tamm-plasmon (TP) modes due to the Purcell-effect in hybrid metal/DBR structures was demonstrated[12,13]. TP states evolving at the interface of a metal layer and a distributed Bragg reflector (DBR) have been theoretically proposed by Kaliteevski et al. in analogy to electronic Tamm surface states[14] and are somewhat similar to optical Fabry-Perot modes, as they can be directly optically excited and are experimentally observed as sharp resonances in the reflection and transmission spectra[15]. The eigenfrequency of such TP states depends both on the thickness of the metal film and the thickness of the adjacent semiconductor layer, allowing to precisely adjust the spectral position of the resonance[14,16]. Lateral confinement of the TP modes can be easily achieved by patterning the metal layer without necessarily etching through the structure, which simplifies device fabrication and makes such structures very attractive in context of scalable production.

Since commercially available low-cost Si-based single photon detectors are most sensitive in the visible red spectral region, it is very favorable to have quantum emitters operating at those wavelengths. In this context QDs in the (In, Ga)P material system are very attractive due to their ability to emit photons in a wide range of the visible red spectrum meeting the aforementioned requirements, and recent experiments have proven their potential to act as SPSs[17,18,19,20]. Finally we have demonstrated site-controlled fabrication of QDs in this material system in previous reports presenting a route towards scalable device integration[21,22].

In this work we present site-controlled InP/GaInP QDs emitting single photons in the visible red spectral range integrated into hybrid metal/dielectric structures and demonstrate the enhancement of their spontaneous emission by coupling to confined TP modes.

Our investigated sample was fabricated on a GaAs (001) substrate by gas-source molecular beam epitaxy. The sample structure is constituted by a DBR made up of 30.5 pairs of AlAs/Al$_{0.40}$Ga$_{0.60}$As and comprises a photonic band gap centered at a wavelength of approximately 660nm at T = 20K. On top of the DBR, a layer of 60nm GaInP lattice matched to GaAs was grown forming the lower barrier. After this epitaxy, square lattices of nanoholes were defined by a combination of optical and electron-beam lithography and an etching technique. InP-QDs were then fabricated in a site-controlled layout on top of an 8nm + 30nm GaInP smoothing and separation layer during overgrowth, which is described in detail elsewhere[21,22]. From an as-grown reference sample with uncapped surface QDs we expect site-controlled QDs to form with a height of around 10-15nm. The QDs were capped by 112nm GaInP, which constitutes the upper barrier. The thickness of this layer was designed to allow for a formation of a TP state in the emission band of the QDs, between approximately 710nm and 650nm[22]. The structure was completed by deposition of a 40nm gold (Au) layer and Au-disks with varying diameters were defined on the sample surface by electron beam lithography and a lift-off technique. Fig. 1 shows the profile of the nominal refractive index throughout the structure together with the intensity of the electric field calculated for a planar structure using the transfer matrix method. The inset shows a calculated reflectivity spectrum assuming a total GaInP thickness of 211nm and a 40 nm thick Au-film with a refractive index for Au at a wavelength around 650nm of $n = 0.147 + i\,3.65$[23]. The reflectivity spectrum shows a prominent dip at 657nm indicating the formation of a TP state.

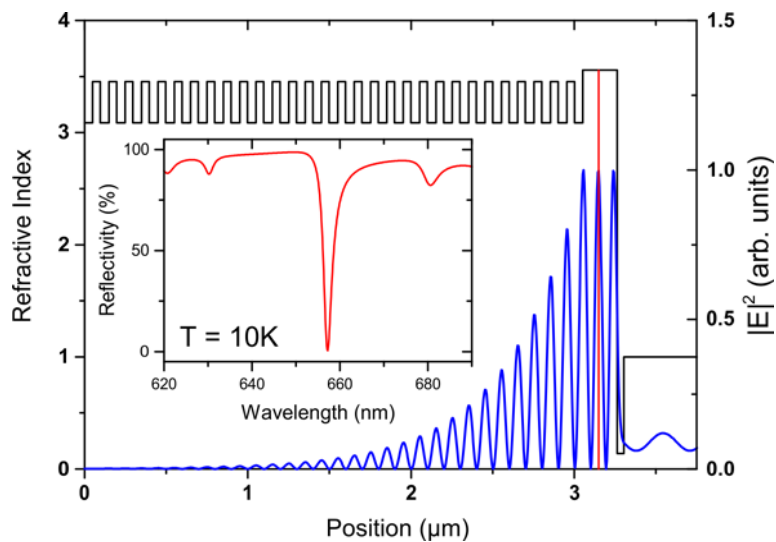

**Fig. 1:** Refractive index profile (black) and calculated profile of the longitudinal electric field intensity (blue) for a planar structure covered with a layer of 40nm Au. The red line indicates the nominal position of the QDs. The inset depicts a calculated reflectivity spectrum at low temperature.

Fig. 2a) presents a scanning electron microscopy image of the sample surface showing the Au-disks with varying diameters ranging from 10µm down to 1µm. For a spatially resolved investigation of the luminescence from our sample, micro-photoluminescence (µPL) spectra at a low temperature of 6K were taken over a total area of 32.5×32.5 µm² every 0.5µm. A frequency-doubled Nd:YAG-laser emitting at a wavelength of 532nm with a power of 750nW was used to excite the QDs. The light was focused through a 20× microscope objective with a

NA of 0.40. The PL signal was analyzed with a 1500 lines/mm reflection grating monochromator and detected by a liquid nitrogen cooled Si charge coupled device camera.

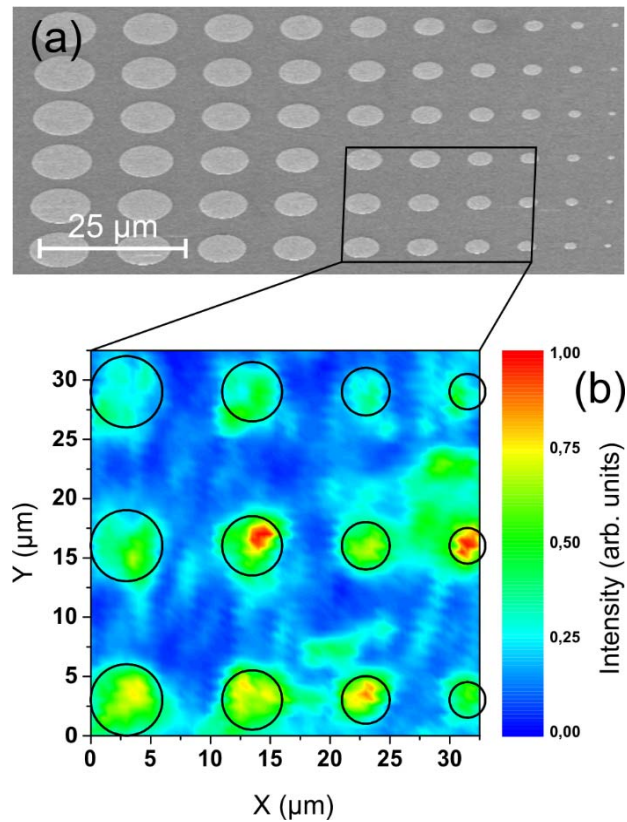

**Fig. 2:** (a) Scanning electron microscopy image of the sample surface showing the array of Au-disks with different diameters ranging from 10μm down to 1μm. (b) Spectrally integrated intensity of a micro-PL area scan in the investigated region with the diameter of the Au-disks ranging from 6μm down to 3μm showing a clearly enhanced PL-intensity of QDs emitting below the Au-disks. The black circles denote the expected position and diameter of the Au-disks.

In Fig. 2b) a micro-photoluminescence (μPL) area scan of the investigated region with disk diameters ranging from 6μm down to 3μm is shown in a false color representation of the luminescence intensity. This luminescence intensity was spectrally integrated in a 6nm window around 655nm corresponding to the wavelength of the TP mode as experimentally observed from photoreflectance measurements on the sample (not shown). In the investigated field the lateral spacing between the predefined nucleation sites for the InP QDs is nominally 500nm (equals an area density of $4\times10^8 cm^{-2}$), so QDs are expected to be located underneath the Au-disks as well as below the surface in between them. From the μPL area scan it can be clearly seen that the luminescence intensity is significantly enhanced at regions associated with the position of the Au-disks. The black circles in Fig. 2a) indicate the expected diameters and the positions of the disks, which very well fit the intensity distribution. The intensities of QDs located underneath an Au-disk are about an order of magnitude higher than for those in bulk material.

A few bright spots can be found in between the Au-discs, which we attribute to particularly bright QDs. A previous investigation on similar site-controlled InP-QDs has shown a strong variation of the luminescence intensity of individual emitters, which is attributed to a variation of their quantum efficiency[24]. The decreasing intensity contrast for the upper disk row may be attributed to a gradual change of the microscope focus during the array scan.

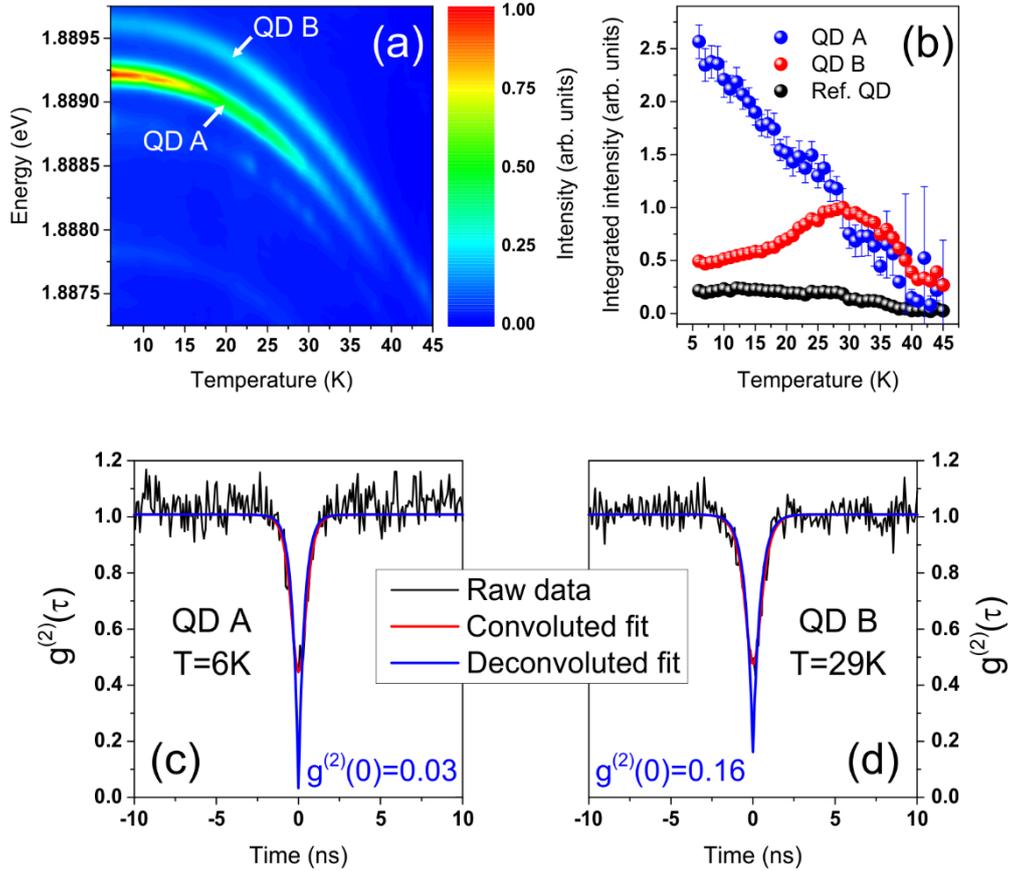

**Fig. 3:** (a) False color representation of micro-PL spectra as a function of the sample temperature, measured on a 4µm Au-disk. The spectra show two prominent features, which we attribute to two QDs: QD A (1.8892eV at T = 6K) and QD B (1.8896eV at T = 6K). (b) Spectrally integrated intensities of the individual emission lines corresponding to QD A and QD B as well as of a reference off-resonance QD as a function of the sample temperature. QD B shows an enhancement of the emission, when tuned in resonance with the confined TP mode, superimposed by thermal intensity quenching. (c), (d) Second-order temporal correlation function $g^{(2)}(\tau)$ of the emission corresponding to QD A (measured at T = 6K) and QD B (measured at T = 29K) in resonance with the confined TP mode respectively.

To investigate the intensity enhancement in more detail we performed temperature tuning experiments of single QDs coupled to a confined TP mode below a disc with a diameter of 4µm. Fig. 3a) depicts µPL spectra at an excitation power of 300nW as a function of the sample temperature in a false color representation of the normalized luminescence intensity. Two dominant emission features can be seen at 6K which we attribute to the emission of two QDs: QD A emitting at 1.8892eV and QD B emitting at 1.8896eV respectively.

Both emission lines show a red-shift with increasing temperature. For the data above T = 25K we can extract Varshni parameters[25] of $\alpha$ = 0.35meV/K and $\beta$ = 240K ($\alpha$ = 0.40meV/K and $\beta$ = 270K) for QD A (QD B) respectively, which indicates that some Ga may be present in the QDs. However the Varshni fit does not account accurately for the low-temperature data and yields large fitting uncertainties. A more accurate fit is achieved by an empirical model assuming electron-phonon coupling[26]. From a fit of this model we can extract an average phonon energy of $\langle\hbar\omega\rangle$ = (7.10±0.10) meV for QD A and $\langle\hbar\omega\rangle$ = (7.37±0.12) meV for QD B. Similar results have been obtained by Schulz et al. for InP QDs in an $(Al_{0.20}Ga_{0.80})_{0.51}In_{0.49}P$ matrix[27]. Temperature tuning is applied to shift the emission lines through the resonance with a Q-factor of approximately 250 estimated from photoreflectance spectra (not shown). Fig. 3b) depicts the evaluated integrated intensities of the emission lines of QD A and QD B as a function of the sample temperature, as well as of a spectrally off-resonant QD (not included in Fig. 3a)) for reference. QD A is already in resonance with the confined TP mode at 6K. With increasing temperature QD A shifts out of resonance and experiences a decrease in intensity, which we believe is a combined effect of reduced coupling to the TP mode and thermal carrier escape. QD B experiences a rise of intensity as the temperature increases with a maximum at 29 K when it is tuned into resonance with the confined TP mode. When this temperature is exceeded, the emission intensity of QD B is reduced again. The weaker emission intensity of QD B which we assume to have a smaller size than QD A is attributed to a rapid quenching of its luminescence due to thermal carrier escape.

In order to prove the non-classical nature of the photon emission into the TP mode we performed temporal photon autocorrelation measurements using a fiber coupled Hanbury Brown and Twiss (HBT) setup. The emission of both QDs was in each case filtered by the monochromator and then focused into the multimode fiber of the HBT setup. The corresponding second order autocorrelation functions $g^{(2)}(\tau)$ are depicted in Fig. 3c) and d) for QD A and QD B respectively. Both reveal a strong antibunching as expected for quantum light emitters. In particular, photon emission with a low probability of multi-photon events is reflected by the observed value of $g^{(2)}(0)_{QD\ A,\ deconv.\ fit}$ = 0.03 ($g^{(2)}(0)_{QD\ A,\ raw\ data}$ = 0.44) for QD A, when taking a finite temporal time resolution of the setup of ~700ps into account. Hereby $g^{(2)}(0)_{raw\ data}$ corresponds to the $g^{(2)}(0)$ value of the raw measurement data and $g^{(2)}(0)_{conv.\ fit}$ is the value extracted by fitting the data points with a Gaussian profile convoluted with a response function. For QD B we obtain $g^{(2)}(0)_{QD\ B,\ deconv.\ fit}$ = 0.16 ($g^{(2)}(0)_{QD\ B,\ raw\ data}$ = 0.48). As the autocorrelation measurement for QD B was performed at an elevated temperature of T = 29K in order to achieve resonance with the TP state, the higher value of $g^{(2)}(0)_{QD\ B}$ is probably related to a smaller signal-to-background ratio due to the weaker intensity of the emission line. For both QDs, the difference of the deconvoluted value of $g^{(2)}(0)$ from the ideal $g^{(2)}(0) = 0$ can be attributed to background emission from the GaInP barrier and scattered light from nearby QDs.

To assess the brightness of our source, we further performed efficiency measurements under pulsed excitation on another QD below a 4µm disk to directly evaluate the performance of our confined TP emitter in terms of the overall efficiency. The µPL spectrum of the investigated QD measured at an excitation power of 1.85µW with the dominant emission peak having a linewidth of 95µeV in resonance with the confined TP mode is depicted in Fig. 4a).

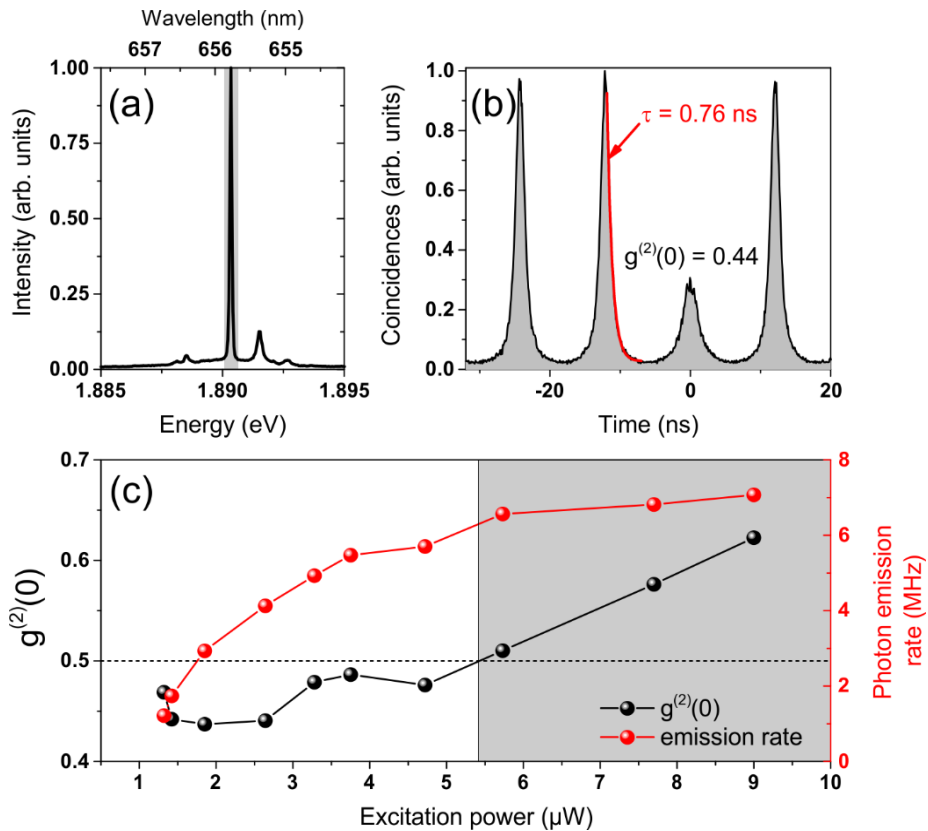

**Fig. 4:** (a) Micro-PL spectrum of a single QD in resonance with the confined TP mode measured on a 4µm Au-disk. The light gray colored region denotes the spectral window, where autocorrelation measurements were performed. (b) Photon autocorrelation measured on the bright emission line shown in (a) under pulsed excitation. The red curve presents a mono-exponential fit to the data yielding a radiative decay time of 0.76ns. (c) True photon emission rate (right axis) and value of the second-order correlation function at $\tau = 0$ (left axis) as a function of the excitation power. The light gray colored background indicates the region where no single photon emission can be observed.

As our 532nm laser source was not capable of delivering short excitation pulses the following experiments were performed using a frequency doubled Ti:sapphire-laser emitting at a wavelength of ~470 nm to excite the QD in the pulsed regime with a repetition rate of 82MHz.

Fig. 4b) shows the result of the autocorrelation measurement in the spectral window indicated by the light gray colored region in Fig. 4a). The extracted $g^{(2)}(0)$ value yields 0.44 for a 2 ns time window and the emission yields a radiative lifetime of 0.76ns determined from the autocorrelation measurement by fitting a mono-exponential decay to the flanks of the peaks.

To determine the photon emission rate $\dot{n}$ we carefully calibrated our setup. To obtain the true single photon emission rate the determined rate $\dot{n}$ has to be corrected for multiphoton emission events: $\dot{n}_{SPE} = \dot{n}\sqrt{1 - g^{(2)}(0)}$. Fig. 4c) displays $\dot{n}_{SPE}$ (red curve) and the corresponding $g^{(2)}(0)$-values (black curve) as a function of the excitation power. For low excitation $\dot{n}_{SPE}$ starts with 1.22 MHz and increases up to 7.07MHz for an excitation power of 9µW. The highest $\dot{n}_{SPE}$-value in the SPS-regime ($g^{(2)}(0)$<0.5) amounts to 5.7MHz at 4.72µW (173k counts per second on the avalanche photodiode (APD)). This value yields an overall SPS efficiency η = 7% when taking the repetition rate of 82MHz into account. This is a distinct enhancement of the efficiency compared to InP/GaInP QDs in bulk material[20] and an improvement by almost a factor of two compared to InP/GaInP QDs in a planar microcavity[28] but nevertheless far below the limit of 60% anticipated for optimized structures[13], which we think is mainly due to absorption losses in the metal layer and interface roughness.

In conclusion we have demonstrated that hybrid metal/semiconductor structures exhibiting the formation of confined TP states can be utilized to enhance the photon emission from InP/GaInP QDs operating in the visible red spectral range. Single photon emission was demonstrated and an SPS efficiency of up to 7% was achieved for QDs in resonance with the confined TP mode. We anticipate that such resonant structures will open a further way besides the well-studied microcavity structures to realize efficient sources of single and entangled-photon sources well compatible with industry-standard Si-based single photon detectors.

This work was financially supported by the EU within the project 'SPANGL4Q' (grant number 284743). S. H. gratefully acknowledges support by the Royal Society and the Wolfson Foundation. Technical assistance by M. Emmerling is gratefully acknowledged.